\documentclass[twocolumn,prl,showpacs,amsfonts,amsmath,assymb,eufrak]{revtex4}

\usepackage{epsfig}

\def\eq#1\en{\begin{equation}#1\end{equation}}  
\def\eqa#1\ena{\begin{align}#1\end{align}}
\def\eqg#1\eng{\begin{gather}#1\end{gather}}
\newcommand{\lb}[1]{\label{e:#1}}
\newcommand{\rlb}[1]{\eqref{e:#1}} 
\newcommand{\nl}{\notag\\}

\newcommand{\sbkt}[1]{\langle#1\rangle}

\newcommand{\sumtwo}[2]%
{\mathop{\sum_{#1}}_{#2}}
\newcommand{\sumthree}[3]%
{\mathop{\mathop{\sum_{#1}}_{#2}}_{#3}}
\newcommand{\sumfour}[4]%
{\mathop{\mathop{\mathop{\sum_{#1}}_{#2}}_{#3}}_{#4}} 
\newcommand{\prodtwo}[2]%
{\mathop{\prod_{#1}}_{#2}}
\newcommand{\mintwo}[2]%
{\mathop{\min_{#1}}_{#2}}
\newcommand{\maxtwo}[2]%
{\mathop{\max_{#1}}_{#2}}
\newcommand{\maxthree}[3]%
{\mathop{\mathop{\max_{#1}}_{#2}}_{#3}}
\newcommand{\limtwo}[2]%
{\mathop{\lim_{#1}}_{#2}}
\newcommand{\suptwo}[2]%
{\mathop{\sup_{#1}}_{#2}}
\newcommand{\supthree}[3]%
{\mathop{\mathop{\sup_{#1}}_{#2}}_{#3}}
\newcommand{\supfour}[4]%
{\mathop{\mathop{\mathop{\sup_{#1}}_{#2}}_{#3}}_{#4}} 
\newcommand{\inftwo}[2]%
{\mathop{\inf_{#1}}_{#2}}
\newcommand{\infthree}[3]%
{\mathop{\mathop{\inf_{#1}}_{#2}}_{#3}}
\newcommand{\inffour}[4]%
{\mathop{\mathop{\mathop{\inf_{#1}}_{#2}}_{#3}}_{#4}} 

\newcommand\calH{{\cal H}}

\newcommand\calU{{\cal U}}

\newcommand{\sfP}{\mathsf{P}}
\newcommand{\sfQ}{\mathsf{Q}}

\newcommand{\sfS}{\mathsf{S}}

\newcommand{\ket}[1]{|#1\rangle}
\newcommand{\psa}{\ket{\psi_\alpha}}
\newcommand{\psb}{\ket{\psi_\beta}}
\newcommand{\ph}{\ket{\varphi}}
\newcommand{\phz}{\ket{\varphi(0)}}
\newcommand{\pht}{\ket{\varphi(t)}}
\newcommand{\kxi}[1]{\ket{\xi^{(#1)}}}
\newcommand{\xij}{\kxi{j}}

\newcommand{\bra}[1]{\langle#1|}

\newcommand{\eoD}{=1,\ldots,D}
\newcommand{\eod}{=1,\ldots,d}
\newcommand{\ab}{{\alpha\beta}}
\newcommand{\suma}{\sum_{\alpha=1}^D}
\newcommand{\sumab}{\sum_{\alpha,\beta=1}^D}
\newcommand{\sumj}{\sum_{j=1}^d}

\newcommand{\Hneq}{\calH_\mathrm{neq}}
\newcommand{\Hrnd}{\calH_\mathrm{rnd}}
\newcommand{\Heq}{\calH_\mathrm{eq}}
\newcommand{\Pneq}{\hat{P}_\mathrm{neq}}
\newcommand{\Prnd}{\hat{P}_\mathrm{rnd}}
\newcommand{\DU}{\mathit{\Delta}U}
\newcommand{\DE}{\mathit{\Delta}E}
\newcommand{\tE}{\tilde{E}}
\newcommand{\hH}{\hat{H}}

\newcommand{\hU}{\hat{U}}

\newcommand{\tauB}{\tau_\mathrm{B}}
\newcommand{\taue}{\tau_\mathrm{esc}}
\newcommand{\tbeta}{\tilde{\beta}}
\newcommand{\tS}{\tilde{S}}
\newcommand{\tsfS}{\tilde{\mathsf{S}}}

\newcommand{\tr}{\mathrm{Tr}}
\newcommand{\Wg}{{\mathrm{Wg}}}
\newcommand{\Sn}{{\mathfrak{S}_n}}

\newcommand{\mumax}{\mu_\mathrm{max}}
\newcommand{\mub}{\bar{\mu}_\mathrm{max}}
\newcommand{\lamax}{\lambda_\mathrm{max}}
\newcommand{\tmax}{\tau_\mathrm{max}}

\newcommand{\lessapprox}
{\mathrel{\raisebox{-2.8pt}{\mbox{$\stackrel{\textstyle <}{\approx}$}}}}

\newcommand{\para}[1]{{\em #1}\/.---}
\newcommand{\midskip}{\vspace{3pt}}

\begin{document}
\title{Extremely quick thermalization in a macroscopic quantum system for a typical nonequilibrium subspace
}

\author{Sheldon Goldstein${}^1$, Takashi Hara${}^2$, and Hal Tasaki${}^3$}
\affiliation{
${}^1$%
Departments of Mathematics and Physics, Rutgers University, 110 Frelinghuysen Road, Piscataway, NJ 08854-8019, USA
\\
${}^2$%
Faculty of Mathematics,
Kyushu University,
Moto-oka, Nishi-ku,
Fukuoka 819-0395,
Japan
\\
${}^3$%
Department of Physics, Gakushuin University, 
Mejiro, Toshima-ku, Tokyo 171-8588, Japan}

\date{\today}

\begin{abstract}
The fact that macroscopic systems approach thermal equilibrium may seem puzzling, for example, because it may  seem to conflict with the time-reversibility of the microscopic dynamics.
We here prove that in a macroscopic quantum system for a typical choice of ``nonequilibrium subspace'', any initial state indeed thermalizes, and in fact does so very quickly, on the order of the Boltzmann time $\tauB:=h/(k_\mathrm{B}T)$.
Therefore what needs to be explained is, not that macroscopic systems approach equilibrium, but that they do so slowly.
\end{abstract}


\maketitle

The recent renewed interest in the foundation of quantum statistical mechanics has led to a revival of the old approach by von Neumann to investigate the problem of thermalization only in terms of quantum dynamics in an isolated system \cite{vonNeumann,GLTZ}.
It has been demonstrated in some general or concrete settings that a pure initial state evolving under quantum dynamics indeed thermalizes (i.e., approaches thermal equilibrium)  in a certain mathematical  sense \cite{Hal1998,Reimann,LindenPopescuShortWinter,
GLMTZ09b,Hal2010,ReimannKastner,Reimann2,SatoKanamotoKaminishiDeguchi}.
See also \cite%
{PopescuShortWinter,GLTZ06,Sugita,SugiuraShimizu12,SugiuraShimizu13}
for closely related idea that a typical pure state of a macroscopic quantum system can fully describe  thermal equilibrium.

Though the above mentioned works establish that a wide class of isolated quantum systems exhibits the approach to equilibrium, they provide basically no information about the necessary time scale.
In our recent work \cite{GHT13}, we have treated this problem of time scale in an abstract setting.
We proved via an explicit (and artificial)
construction that, depending on the system (or,
more precisely, the choice of nonequilibrium subspace $\Hneq$ introduced later), the
relaxation time can be extremely large, exceeding the age
of the universe, or ridiculously short.

This motivates us to study the time scale of thermalization, or, more precisely, that of the escape from the nonequilibrium subspace $\Hneq$, for various choices of $\Hneq$.
It may be natural to first focus on a setting in
which the ``nonequilibrium subspace'' $\Hneq$ is not, in fact, a realistic nonequilibrium subspace, but rather is chosen in a
completely random manner.
One might hope that for such a subspace one
 generically has a realistic relaxation time.
If this were so, it would seem reasonable to believe that the same thing would probably be true for a realistic nonequilibrium subspace (unless we have reasons to expect otherwise).
This is basically von Neumann's philosophy in \cite{vonNeumann}.
Unfortunately this hope turns out to be far too
optimistic.

In the present paper, we report on our recent mathematically
rigorous (and probably rather surprising) result
implying that, for {\em any}\/ initial state (and hence for those that are far from equilibrium), the relaxation time in the above
setting is quite short, of order the
Boltzmann time $\tauB:=h/(k_\mathrm{B}T)$.
Since $\tauB\approx10^{-13}\rm\,s$ at room temperature, we have basically proved that our coffee is no longer hot after, say, a micro-second.
This physically absurd
but mathematically rigorous conclusion has the important
implication that the moderately slow decay observed in
reality is {\em not}\/ a typical property for a random
choice of $\Hneq$.
It suggests that we have to take
into account some essential features of realistic nonequilibrium
subspaces in order to fully understand the problem
of thermalization in isolated quantum systems.
We shall discuss possible directions for future research.

In the present paper, we discuss our main result, its physical implications, and rough ideas of the proof.
Full mathematical details can be found in \cite{GHT13full}.

\midskip
\para{Setting and main result}
Our setting is basically the same as in our previous work \cite{GHT13}.
Following von Neumann \cite{vonNeumann,GLTZ} and Goldstein, Lebowitz, Mastrodonato, Tumulka, and Zangh\`\i\ \cite{GLMTZ09b}, we consider an abstract model of an isolated macroscopic quantum system in a large volume. 
A typical example is a system of $N$ particles confined in a box of volume $V$, where the density $N/V$ is kept constant when $V$ becomes large.
In what follows we assume that the volume $V$ is fixed, and do not discuss the $V$ dependence of various quantities explicitly.

Let $\hH$ be the Hamiltonian, and denote by $E_\alpha$ and $\psa$ the eigenvalue and  the normalized eigenstate, respectively, of $\hH$, i.e.,
$\hH\psa=E_\alpha\psa$.
We focus on the energy interval $[U-\DU,U]$, where $\DU$ denotes the range of energy, which is small from the macroscopic point of view, but is still large enough to contain many energy levels.
We also assume $\DU\gg k_\mathrm{B}T$, where $T$ is defined below the theorem.
The choice of $\DU$ is somewhat arbitrary.
We relabel the index $\alpha$ so that the energy eigenvalues $E_\alpha\in[U-\DU, U]$  precisely correspond to the indices $\alpha\eoD$.
We shall work with the Hilbert space $\calH$ spanned by all $\psa$ with $\alpha\eoD$, which is often called the {\em microcanonical energy shell}\/.

We decompose $\calH$ into the equilibrium and the nonequilibrium subspaces as $\calH=\Heq\oplus\Hneq$.
Any state $\ph$ which is close enough to $\Heq$ represents
the equilibrium state, and a state not close to $\Heq$ represents
a nonequilibrium state. 
We further assume that
the dimension $d$ of $\Hneq$ satisfies $1\le d\ll D$ \cite{en:d}. 
This corresponds to the fact that the overwhelming majority of states in $\calH$ correspond to the thermal equilibrium state
\cite{GLTZ,GLMTZ09b,Hal2010,PopescuShortWinter,GLTZ06,Sugita}.

We then take an arbitrary normalized initial state $\phz\in\calH$, which may be close to $\Hneq$, and consider its time-evolution $\pht=e^{-i\hH t/\hbar}\phz$.
Let $\Pneq$ be the projection onto $\Hneq$.
If it happens for some $\tau>0$ that
\eq
\frac{1}{\tau}\int_0^\tau dt\,\bra{\varphi(t)}\Pneq\pht\ll1,
\lb{approach}
\en
we can say that,  within the interval $[0,\tau]$, the state $\pht$ escapes from the vicinity of $\Hneq$, or comes close to $\Heq$ and stays there for most time.
Thus the system thermalizes within $\tau$.

In reality the subspace $\Hneq$ should be almost uniquely determined from physical properties of the system.
It is not an easy task, however, to characterize $\Hneq$ in general or in a concrete setting. 
Nor is it easy to usefully estimate the relaxation time for any specific physical $\Hneq$. 
We shall therefore take an abstract approach in which we regard $\Hneq$ as a general $d$-dimensional subspace of $\calH$, and try to elucidate the relation between $\Hneq$ and the associated relaxation time.

Our finding in the previous work \cite{GHT13} is that, for certain choices of $\Hneq$, the required time $\tau$  becomes as large as $d\,\hbar/\DU$ or as small as $\hbar/\DU$.

In the present paper we shall focus on the relaxation time associated with a generic subspace $\Hneq$.
For this purpose we choose the subspace in a completely random manner as follows.
Let $\calU(\calH)$ be the group of all 
unitary transformations on $\calH$.
For each $\hU\in\calU(\calH)$ and $j\eod$, we define a normalized state $\xij=\hU\ket{\psi_j}$.
We then consider the $d$-dimensional subspace $\Hrnd$ spanned by $\kxi{1},\ldots,\kxi{d}$, and the corresponding projection operator 
\eq
\Prnd=\sumj\kxi{j}\bra{\xi^{(j)}}.
\lb{Pdef}
\en
By drawing $\hU\in\calU(\calH)$ according to the unique Haar measure on $\calU(\calH)$, we can generate the $d$-dimensional subspace $\Hrnd$ and the associated projection $\Prnd$ in a completely random manner.
The random subspace $\Hrnd$ is our probabilistic model for $\Hneq$.

Then our main result, implying that relaxation typically can't take much longer than the Boltzmann time $\tauB$ (see below), is the following.
\par\noindent
\para{Theorem}
Let $1\le d\ll D$.
With probability close to one, we have \cite{en:approx}
\eq
\frac{1}{\tau}\int_0^\tau dt\,\bra{\varphi(t)}\Prnd\pht
\lesssim\frac{\tauB}{\tau},
\lb{main}
\en
for any normalized $\phz\in\calH$ and any $\tau$ such that $0<\tau\le\tmax$, where $\tmax:=\tauB\,\min\{(D/d)^{1/4},D^{1/6}\}$ is a large constant \cite{en:longtau}.

Mathematically precise assumptions and statements can be found in \cite{GHT13full}.

In \rlb{main}, we have introduced the Boltzmann time $\tauB:=h/(k_\mathrm{B}T)$, where the inverse temperature $T$ is determined by the standard formula $(k_\mathrm{B}T)^{-1}=\partial\log\Omega(E)/\partial E|_{E=U}$ involving the number of states $\Omega(E)$.
We assume that $\Omega(E)$ exhibits the normal behavior $\Omega(E)\approx\exp[V\,\sigma(E/V)]$, where the entropy density $\sigma(\epsilon)$ is an increasing function \cite{Ruelle}.
For our theorem to be valid, it is essential to take into account properly the behavior of $\Omega(E)$.

\midskip
\para{Discussions}
We first note that the Boltzmann time is $\tauB\sim1.6\times10^{-13}\rm\,s$ at $T\sim300\rm\,K$.
Then even for $\tau\approx10^{-6}\rm\,s$, the right-hand side of \rlb{main} does not exceed $10^{-6}$.
This means that the system, starting from {\em any}\/ initial state, equilibrates much before one micro-second.

Although the Boltzmann time $\tauB$ is an important time scale for many quantum phenomena, it is simply absurd that a macroscopic system always thermalizes so quickly. 
We also note that the bound \rlb{main} reflects no information about the size of the system.
This is unphysical since, in general, the relaxation time should increase with the size of the system.
Since the theory is rigorous, the only reasonable conclusion is that realistic nonequilibrium subspaces form exceptions to the bound of the theorem, or, in other words, the moderately slow decay observed in reality is not typical (if we assume random $\Hneq$) \cite{anotherInterpretation}.

The atypicality may not be too surprising, especially after knowing about it.
Given the energy shell $\calH$, the nonequilibrium subspace $\Hneq$, in reality, is determined not in a random manner, but through the values of macroscopic quantities that we use to characterize the system.
Recall that many of the standard macroscopic quantities are expressed as a sum (or an integral) of locally conserved observables, and the Hamiltonian of a realistic system consists of more or less short-range interactions.
This means that the corresponding subspace $\Hneq$ and the projection 
operator $\Pneq$ should be special.
It is likely, for example, that the commutator $[\hH,\Pneq]$ is smaller for realistic subspaces compared with randomly chosen ones. 

In order to fully understand the problem of the approach to equilibrium in macroscopic quantum systems, it may be essential to characterize realistic nonequilibrium subspaces, and to investigate the accompanying time scale.

The escape from a single state may provide a hint for such considerations.
Take an arbitrary state $\ket{\xi}\in\calH$ such that $\sbkt{\psi_\alpha|\xi}$ is negligible (in a certain rough sense) unless $\alpha$ is such that $E_\alpha\in[\tE,\tE+\DE]\subset[U-\DU,U]$ for some energy $\tE$ and energy width $\DE$.
Then take an initial state $\phz\in\calH$ which is close to $\ket{\xi}$, and examine how quickly the state escapes from the vicinity of $\ket{\xi}$.
As is well-known as one form of the ``uncertainty relation between time and energy'', the overlap $\bigl|\sbkt{\xi|\varphi(t)}\bigr|^2$ changes (and hence decays) in the time scale of order $\taue:=h/\DE$.

This observation suggests that a subspace $\Hneq$ may also be characterized by certain energy scale $\DE$, which is similarly related to the associated relaxation time.
This picture is true, at least for some examples, as we shall see now.

In the present setting of completely random $\Hneq$, we see that each $\kxi{j}$ is characterized by the width (in the above sense) $\DE\approx k_\mathrm{B}T$.
This is  because, in a macroscopic system, most of the energy eigenvalues $E$ such that $U-\DU\le E\le U$ are found in the smaller range
with $U-\text{const}\,k_\mathrm{B}T\le E\le U$, where the constant is of $O(V^0)$. 
From this the conclusion that  the escape time coincides with the Boltzmann time does indeed  follow.

Recall that we made a reasonable assumption that $\DU\gg k_\mathrm{B}T$.
In the unphysical case with $\DU\ll k_\mathrm{B}T$, our inequality \rlb{main} should be modified and the right-hand side becomes $h/(\DU\,\tau)$.
This is again consistent with our picture since each $\kxi{j}$ now has width $\DU$, which corresponds to the escape time $\taue=h/\DU$.

Two examples of $\Hneq$ in our previous work \cite{GHT13} are also consistent with the picture of escape time.
In Theorem~1, where one finds extremely slow decay, we have $\DE=\DU/d$, which means $\taue=h\,d/\DU$.
In Theorem~2, where one finds quick decay, we have $\DE\sim\DU$, which means $\taue\sim h/\DU$.

This observation suggests that 
a realistic nonequilibrium
subspace $\Hneq$, determined through macroscopic quantities
that we use to characterize the system, is associated with a certain energy width $\DE$.
It is possible that the escape time $\taue$, which may take a reasonable value depending on the value of $\DE$, essentially determines the relaxation time.

Of course it is also likely that the above picture of the escape from a single state is too naive or has only limited applicability.
In some systems it may happen that $\pht$ is ``trapped'' in the vicinity of $\Hneq$ in a more intricate manner.
In such a situation the relaxation time should also depend on the number of independent $\kxi{j}$'s that the state $\pht$ should go through.

It is certainly an interesting challenge to examine these pictures in interacting many-body quantum systems.
For the moment we only have limited rigorous results in abstract and artificial settings.
In particular Theorem~1 of \cite{GHT13} treats examples with unphysically long relaxation time, and the theorem of the present paper shows that a typical $\Hneq$ leads to unphysically short relaxation time.
The reality should lie in between these two extreme theorems, remaining to be understood. 

\midskip
\para{Sketch of proof}
Since a mathematically complete
proof can be found in \cite{GHT13full}, we here briefly describe the essential ideas.

Let $\phz=\suma c_\alpha\psa$ be an arbitrary normalized initial state.
Then we write
\eq
\frac{1}{\tau}\int_0^\tau dt\,\bra{\varphi(t)}\Prnd\pht
=\sumab(c_\alpha)^*Q_\ab\,c_\beta,
\lb{PcQc}
\en
where the matrix $\sfQ=(Q_\ab)$ is defined by
\eq
Q_\ab:=S_\ab\,P_\ab,
\lb{Q=PS}
\en
with $P_\ab:=\bra{\psi_\alpha}\Prnd\psb$, and
\eq
S_\ab:=
\frac{1}{\tau}\int_0^\tau dt\,e^{i(E_\alpha-E_\beta)t/\hbar}.
\lb{Sdef}
\en
The matrix $\sfQ$ is the Hadamard product of the matrices $\sfP=(P_\ab)$ and $\sfS=(S_\ab)$.
Note also that $\sfQ$ is hermitian since both $\sfP$ and $\sfS$ are hermitian.

Let $\mumax$ be the maximum eigenvalue of $\sfS$.
When $\tau$ is not too large, we have the following bound for $\mumax$.
\par\noindent
\para{Proposition 1}
If $D\gg1$ and $0<\tau\le\tmax$, we have $\mumax\lesssim D\,\tauB/\tau$.

Let $\lamax$ be the maximum eigenvalues of $\sfQ$.
The following proposition, which bounds $\lamax$ by a bound on $\mumax$, is our main mathematical result.
\par\noindent
\para{Proposition 2}
Let $1\le d\ll D$.
Suppose that there is a constant $\mub$ such that $\mumax\le\mub$ and $\mub\gg d$.
Then we have $\lamax\lesssim \mub/D$ with probability close to one.

When $0<\tau\le\tmax$, Proposition~1 allows us to choose $\mub\simeq D\,\tauB/\tau$, which satisfies $\mub\gg d$.
By noting that \rlb{PcQc} implies
\eq
\frac{1}{\tau}\int_0^\tau dt\,\bra{\varphi(t)}\Prnd\pht
\le\lamax,
\lb{Plamax}
\en
we get the desired theorem from Proposition~2.

\midskip
\para{On Proposition 1}
Since the proof in \cite{GHT13full} of the upper
bound is rather technical, let us here give a simple variational argument (which indeed proves the corresponding {\em lower}\/ bound)
which sheds light on the nature of $\mumax$ and the corresponding eigenvector.

It is convenient to define the matrix $\tsfS=(\tS_\ab)$ by
\eq
\tS_\ab:=
\frac{1}{\tau}\int_{-\tau/2}^{\tau/2}dt\,e^{i(E_\alpha-E_\beta)t/\hbar}.
\lb{tSdef}
\en
The matrix $\tsfS$  has exactly the same eigenvalues as $\sfS$, and is real symmetric.
Let us approximate the density of states as $\rho(E)\sim D\tbeta\,e^{\tbeta(E-U)}$, where $\tbeta=(k_\mathrm{B}T)^{-1}$ is the inverse temperature. 
We have extended the range of energy to $-\infty<E\le U$.
The approximation is reasonable in a macroscopic system.

Since $\tS_\ab$ has oscillating signs, the eigenvector corresponding to $\mumax$ should have nonvanishing components concentrated in a small energy range with high density of states.
This motivates us to choose a normalized trial state $\varphi_\alpha=f(E_\alpha)$ with $f(E):=A\,e^{\gamma(E-U)}$.
The normalization constant is $A\sim\sqrt{(2\gamma+\tbeta)/(D\tbeta)}$.
Then the expectation value is easily computed as
\eqa
&\sumab\varphi_\alpha\tS_\ab\varphi_\beta
\sim\frac{1}{\tau}\int_{-\tau/2}^{\tau/2}dt\int_{-\infty}^UdE\,dE'
\nl&
\hspace{29mm}\times\rho(E)\,\rho(E')\,f(E)\,f(E')\,e^{i(E-E')t/\hbar}
\nl&
\hspace{20mm}=D\,\frac{\tbeta\hbar}{\tau}\,
\frac{2\gamma+\tbeta}{\gamma+\tbeta}\,
2\arctan\Bigl[\frac{\tau}{2(\gamma+\tbeta)\hbar}\Bigr],
\ena
which becomes $D\,\tauB/\tau$ when $\tau/\hbar\gg\gamma\gg\tbeta$.

\midskip
\para{On Proposition 2}
It can be shown that $\sfQ$ is positive semi-definite.
Thus we have $\tr[\sfQ^n]\ge(\lamax)^n$.
Then note that
\eqa
&\overline{\tr[\sfQ^n]}
=\sum_{\alpha_1, \ldots, \alpha_n=1}^D
\overline {
	Q_{\alpha_1 \alpha_2} \, Q_{\alpha_2 \alpha_3} \, \cdots \,  
	Q_{\alpha_n \alpha_1} }
	\nl
	& = \sum_{\alpha_1, \ldots, \alpha_n=1}^D 
	S_{\alpha_1\alpha_2} \, \cdots \,  
	S_{\alpha_n \alpha_1} \, 
	\overline{
	P_{\alpha_1 \alpha_2}  \, \cdots \,  
	P_{\alpha_n \alpha_1} 
	},
	\lb{TrBn1}
\ena
where the bar denotes the average over the random choice of $\kxi{j}$'s.
The definition \rlb{Pdef} implies 
$P_\ab=\bra{\psi_\alpha}\Prnd\psb=
\sum_{j=1}^d\sbkt{\psi_\alpha|\xi^{(j)}}\sbkt{\xi^{(j)}|\psi_\beta}=
\sum_{j=1}^d\xi^{(j)}_\alpha(\xi^{(j)}_\beta)^*$, 
where we wrote $\xi^{(j)}_\alpha:=\sbkt{\psi_\alpha|\xi^{(j)}}$.
Then the expectation in \rlb{TrBn1} is written as
\eqa
&\overline{
	P_{\alpha_1 \alpha_2}  \, P_{\alpha_2 \alpha_3}\,\cdots \,  
	P_{\alpha_n \alpha_1} 
	}	\nl&
	= \sum_{j_1, \ldots, j_n=1}^d
	\overline{
	\xi^{(j_1)}_{\alpha_1} \,(\xi^{(j_1)}_{\alpha_2})^* \, 
	\cdots 
	\xi^{(j_n)}_{\alpha_{n}} \,
	(\xi^{(j_n)}_{\alpha_{1}})^* \, 
	}
	.
	\lb{aveP.11}
\ena
In a crude approximation  we treat $\kxi{j_1},\ldots,\kxi{j_n}$ as independent random vectors each satisfying the normalization condition 
$\overline{\xi^{(j)}_{\alpha} \, (\xi^{(j)}_{\alpha'})^*}
=D^{-1}\,\delta_{\alpha,\alpha'}$.
This seems reasonable provided that $n\ll d\ll D$ \cite{en:d>>1}, and was partially justified by Weingarten \cite{Weingarten}.

Then the relevant expectation becomes
\eq
	\overline{
	\xi^{(j_1)}_{\alpha_1} \,(\xi^{(j_1)}_{\alpha_2})^* \, 
	\xi^{(j_2)}_{\alpha_2} \,
	\cdots 
	\xi^{(j_n)}_{\alpha_{n}} \,
	(\xi^{(j_n)}_{\alpha_{1}})^* \, 
	}
	\sim
	\frac{1}{D^n}\,\prod_{s=1}^n\delta_{\alpha_s,\alpha_{s+1}},
	\lb{free2}
\en
where we identified $\alpha_{n+1}$ with $\alpha_1$.
Substituted into \rlb{aveP.11}, this approximation yields
\eq
	\overline{
	P_{\alpha_1 \alpha_2} \, P_{\alpha_2 \alpha_3} \, \cdots \,  
	P_{\alpha_n \alpha_1} 
	}
	\sim \Bigl(\frac{d}{D}\Bigr)^n\,
	\prod_{s=1}^n\delta_{\alpha_s,\alpha_{s+1}}.
	\lb{free3}
\en
Note that the factor $\prod_{s=1}^n\delta_{\alpha_s,\alpha_{s+1}}$ imposes the constraint that $\alpha_1,\ldots,\alpha_n$ must be all identical.
Thus, recalling \rlb{TrBn1}, the present approximation gives
\eq
\overline{\tr[\sfQ^n]}\stackrel{\text{?}}{\approx}
\Bigl(\frac{d}{D}\Bigr)^n\sum_{\alpha=1}^D(S_{\alpha\alpha})^n
=D\Bigl(\frac{d}{D}\Bigr)^n.
\lb{free4}
\en
where we noted that $S_{\alpha\alpha}=1$ by \rlb{Sdef}.
This cannot be the main contribution as it is too small and, moreover, independent of $\tau$.

We next set $j_1=\cdots=j_n=j$ in the expectation value
	$\overline{
	\xi^{(j_1)}_{\alpha_1} \,(\xi^{(j_1)}_{\alpha_2})^* \, 
	\cdots 
	\xi^{(j_n)}_{\alpha_{n}} \,
	(\xi^{(j_n)}_{\alpha_{1}})^* \, 
	}$
to get
\eqa
	&\overline{
	\xi^{(j)}_{\alpha_1} \,(\xi^{(j)}_{\alpha_2})^* \, 
	\xi^{(j)}_{\alpha_2} \,(\xi^{(j)}_{\alpha_3})^* \, 
	\cdots 
	\xi^{(j)}_{\alpha_{n}} \,
	(\xi^{(j)}_{\alpha_{1}})^*\,
	}
	\nl&
	=
	\overline{
	|\xi^{(j)}_{\alpha_1}|^2\,|\xi^{(j)}_{\alpha_2}|^2
	\cdots|\xi^{(j)}_{\alpha_n}|^2
	}
	\sim\frac{1}{D^n},
	\lb{free6}
\ena
where we assumed for simplicity that all $\alpha_1,\ldots,\alpha_n$ are distinct, and used the fact that for any $j$ the coefficients $\xi_\alpha^{(j)}$ (with $\alpha=1,\ldots,D$) of the random vector $\xij$ can be treated as independent random variables.
Assuming that \rlb{free2} and \rlb{free6} give the dominant contributions, we find from \rlb{aveP.11} that
\eq
	\overline{  
	P_{\alpha_1 \alpha_2} \, P_{\alpha_2 \alpha_3} \, \cdots \,  
	P_{\alpha_n \alpha_1} 
	}
	\approx \Bigl(\frac{d}{D}\Bigr)^n\,
	\prod_{s=1}^n\delta_{\alpha_s,\alpha_{s+1}}+\frac{d}{D^n}.
	\lb{free7}
\en
Note that the first term in the right-hand side, which is \rlb{free3}, is larger
but has the constraint on the  $\alpha$'s while the second term is smaller
but is (almost) free from constraint.
Going back to \rlb{TrBn1}, the second term yields a new contribution
\eq
	\sum_{\alpha_1, \ldots, \alpha_n=1}^D 
	S_{\alpha_1\alpha_2} \, S_{\alpha_2 \alpha_3} \, \cdots \,  
	S_{\alpha_n \alpha_1} \, 
	\frac{d}{D^n}
	=\tr[\sfS^n]\,\frac{d}{D^n}.
	\lb{free8}
\en
Since $\sfS$ is easily found to be positive semi-definite, we have $\tr[\sfS^n]\le D\,(\mumax)^n\le D\,(\mub)^n$.
We thus arrive at the estimate
\eq
\frac{1}{D}\,\overline{(\lamax)^n}\lessapprox\Bigl(\frac{d}{D}\Bigr)^n+\Bigl(\frac{\mub}{D}\Bigr)^n\,d
\sim \Bigl(\frac{\mub}{D}\Bigr)^n\,d.
\lb{free9}
\en
By taking $n$ sufficiently large, Proposition 2 then follows from the Markov inequality.

Since $\kxi{j}$'s are in fact correlated with each other, it is a nontrivial task to make the above estimate into a rigorous one.
Fortunately we can make use of the recent exact integration formula due to Collins \cite{Coll03,CS06}, which gives
\eqa
	&\overline{
	\xi^{(j_1)}_{\alpha_1} \,(\xi^{(j_1)}_{\alpha_2})^* \, 
	\xi^{(j_2)}_{\alpha_2} \,(\xi^{(j_2)}_{\alpha_3})^* \, 
	\cdots 
	\xi^{(j_n)}_{\alpha_{n}} \,
	(\xi^{(j_n)}_{\alpha_{1}})^* \, 
	}
	\nl&
	= \sum_{\sigma, \tau \in \Sn} 
	I[\forall k,  \alpha_{k} = \alpha_{\sigma(k)+1},\,
	j_k = j_{\tau(k)}  ] \, 
	\Wg(\tau \sigma^{-1}),
	\lb{Avxi=IW1}
\ena
where $\Sn$ is the group of all permutations of $\{1,2,\ldots,n\}$.  
The Weingarten-Collins function $\Wg(\sigma)$ behaves for large $D$ as
$|\Wg(\sigma)|\approx 1/D^{n+|\sigma|}$, where $|\sigma|$ denotes the minimum number of transpositions necessary to express the permutation $\sigma$ as their products.
By making use of the formula \rlb{Avxi=IW1} and some properties of the group $\Sn$, we can control all the terms which contribute to $\overline{\tr[\sfQ^n]}$, and prove the theorem \cite{GHT13full}.

\bigskip

We wish to thank
Tetsuo Deguchi,
Takaaki Monnai,
Shin-ichi Sasa,
Akira Shimizu,
Ayumu Sugita, 
Sho Sugiura,
and
Yu Watanabe
for valuable discussions and comments.
We also thank Beno\^{i}t Collins, Hiroyuki Ochiai and Tomoyuki Shirai for their help in  mathematical issues.

The present work was  supported in part by grant no.~37433 from the John Templeton Foundation (S.G.), JSPS Grants-in-Aid for Scientific Research nos.~25610021 (T.H.) and 25400407 (H.T.).



\begin{thebibliography}{10}
\bibitem{en:approx}
Throughout the present paper, $A\sim B$ means that $A/B$ is close to one.
$A\approx B$ means the weaker relation that $A/B$ is $O(1)$.
Likewise, $A\lesssim B$ and $A\lessapprox B$ mean $A\le B'$ with $B'\sim B$ and $B'\approx B$, respectively.
\bibitem{en:longtau}
If $\tau/\tauB$ is much larger, the right-hand side of \rlb{main} should be replaced by $d/D$.
\bibitem{anotherInterpretation}
One may interpret our theorem as providing information about how quickly states vary in a macroscopic quantum system.
We find that, with probability close to one, any state (including mixed states) in $\Hrnd$ escapes from $\Hrnd$ on the order of the Boltzmann time.
\bibitem{en:d>>1}
It is sufficient to prove Proposition~2 for large enough $d$, since $\lamax$ is easily found to be non-decreasing in $d$ \cite{GHT13full}.
\end{thebibliography}
\end{document}